# Using Bluetooth Low Energy (BLE) Signal Strength Estimation to Facilitate Contact Tracing for COVID-19

Gary F. Hatke[1] (Hatke@ll.mit.edu), Monica Montanari[1], Swaroop Appadwedula[1], Michael Wentz[1], John Meklenburg[1], Louise Ivers[2], Jennifer Watson[1], Paul Fiore[1]

MIT Lincoln Laboratory, 244 Wood St, Lexington MA 02420

**Abstract:**

The process of contact tracing to reduce the spread of highly infectious and life-threatening diseases has traditionally been a primarily manual process managed by public health entities. This process becomes challenged when faced with a pandemic of the proportions of SARS-CoV2. Digital contact tracing has been proposed as way to augment manual contact tracing and lends itself to widely proliferated devices such as cell phones and wearables. This paper describes a method and analysis of determining whether two cell phones, carried by humans, were in persistent contact of no more than 6 feet over 15 minutes using Bluetooth Low Energy signals. The paper describes the approach to detecting these signals, as well as a data-driven performance analysis showing that larger numbers of samples with more optimal detection algorithms, coupled with privacy preserving auxiliary information, improves detection performance.

## Introduction

COVID-19, the disease which occurs when infected with the SARS-CoV2 virus, is primarily propagated through the population via close contact between an infected person ("index case") and another person. The virus may be spread prior to the appearance of symptoms in the infected individual [1,2]. To stem the spread of the disease, minimizing contact between potentially infected persons and other non-infected persons is required. Currently public health entities engage in a triad of contact tracing, testing, and quarantine to minimize the spread. The current approach to contact tracing is a primarily manual approach which relies on interviews with infected individuals [3,4].

Identifying these close contacts (which, per current CDC guidelines, are those who have spent over 15 minutes within 6 feet of an index case [4-6]) may be facilitated by using automated tracing apps deployed on cell phones. These apps continuously emit Bluetooth low energy (BLE) "chirps" which other phones will periodically scan, and record the pseudo-random bit sequence emitted by the chirping phone as well as the estimated power with which the chirp was received [7-9].

The PACT [3] [10] program has proposed a method for making a binary decision based on received BLE signal power estimates: are the two phones "Too Close For Too Long?" (TCFTL). Previous work has looked at the performance of some limited detector strategies using BLE power measurement as a discriminator for relative range between phones [11]. Our work looks at optimizing detector strategies as a function of available auxiliary data and additional BLE power measurements which may be available to the phone.

A key factor in declaring that a contact is TCFTL is determining that the range between the phones is less than 6 feet. Unfortunately, accurately determining

DISTRIBUTION STATEMENT A. Approved for public release. Distribution is unlimited.

This material is based upon work supported by the Defense Advanced Research Projects Agency under Air Force Contract No. FA8702-15-D-0001. Any opinions, findings, conclusions or recommendations expressed in this material are those of the author(s) and do not necessarily reflect the views of the Defense Advanced Research Projects Agency.

[1] MIT Lincoln Laboratory, 244 Wood St, Lexington MA 02420

[2] Massachusetts General Hospital Center for Global Health and Harvard Medical School.

[3] PACT is a collaboration led by the MIT Computer Science and Artificial Intelligence Laboratory (CSAIL), MIT Internet Policy Research Initiative, Massachusetts General Hospital Center for Global Health and MIT Lincoln Laboratory. It includes close collaborators from a number of public and private research and development centers and is a partnership among cryptographers, physicians, privacy experts, scientists and engineers. PACT's mission is to enhance contact tracing in pandemic response by designing exposure detection functions in personal digital communication devices that

range between RF devices simply by measuring received power requires knowing many system parameters a-priori, including transmit power into the antenna, both transmit and receive antenna patterns and relative orientations, multipath environment, and shadowing between the devices. In a TCFTL test scenario, other than the raw power into the transmit antenna, none of these are likely to be known. This makes estimating the "too close" portion of TCFTL quite difficult.

We propose to solve the problem by leveraging two factors: knowledge of the relative carriage of the transmitting and receiving devices (such as if the device is being hand-carried or residing in a pocket), and the fact that over the "too long" period, we will have multiple measurements of received signal power. This paper describes how we leverage this information, including data collection campaigns to understand some of the typical propagation variations seen in a TCFTL scenario. We then evaluate the performance of various detectors, and make predictions on overall performance of the TCFTL detector in a contact tracing framework.

**BLE Signal Power Phenomenology**

BLE signaling in the context of COVID-19 contact tracing is done through "advertisements", which are short messages (typically less than 500 bits) at relatively high data rates (~1 Mb/s) that are repeatedly broadcast from a device at roughly 250 ms intervals. These advertisements can contain limited data, including the pseudo-random bit sequence generated by the transmitting device, as well as some indication of the status of the transmitting phone. The transmissions occur on a subset of three of the BLE frequency channels (2402 MHz, 2426 MHz, and 2480 MHz), roughly occurring at the low, mid, and high frequency ranges of the BLE allotted frequencies [12-14]. In the COVID-19 contact tracing implementation, the transmitted signal power into the antenna is encoded into the advertisement, allowing the receiver to estimate what power level was transmitted by the advertising phone, which in turn allows the receiving phone to produce an estimate of the path loss (attenuation) of the signal from the transmitting device to the receiving device.

The receiver, as implemented on Android and iOS devices, will make an estimate of the received signal power, referred to as an RSSI (Received Signal Strength Indication) measurement, which is nominally the Decibel representation of received power in milliwatts (dBm). This estimate is quantized in 1 dB increments. Most devices have sensitivities which allow demodulation of packets with RSSI values of -100 dBm or greater.

It is well known that in the 2.4 GHz ISM band where BLE signals exist, signal attenuation due to human shadowing (a human in close proximity to either the transmitting or receiving device, blocking direct line of sight propagation) can be quite large. In addition, the antenna patterns of BLE devices such as smart phones can be highly anisotropic, with large variations in gain as a function of angle and polarization. Finally, propagation in the 2.4 GHz ISM band in indoor environments is subject to potentially deep multipath (Rayleigh) fading from nearby reflections. All of these factors make the job of determining range through only power measurements quite difficult [15].

To understand the severity of these variable conditions, we undertook a measurement campaign using actual iOS devices equipped with specially designed "apps" which allowed for rapid collection of RSSI data between pairs of phones which were a known distance apart [16]. Because we needed to understand the variations in attenuation as a function of personal carriage of the phones, as well as the relative position of the phone carriers with respect to the line of sight between the two phones, we took repeated measurements with phones in different carriage positions (front pants pocket, back pants pocket, shirt pocket, in hand, in bag), standing vs. sitting [17], and with different rotations of the two individuals with respect to the line of sight direction (measurements were made every 45 degrees of rotation). For each rotational pose angle, separation distance, and phone carriage state, multiple RSSI measurements were made. This was important due to the fact that each RSSI measurement is made on a single BLE frequency – and the Rayleigh scattering will typically be frequency dependent.

Because these measurements were made during the period of time where strict social distancing was in place, all measurements were conducted in private homes, using family members for phone carriage. To speed up the process, only one phone holder rotated their pose during a measurement (user 1), producing 8 angular measurements per phone separation distance. To emulate the rotation of the carrier of the stationary phone (user 2), we modeled the effect of pose variation as a bulk delta in attenuation which was dependent on pose angle and phone carriage state. These bulk attenuation parameters were estimated in data sets where a given phone carriage state was in use by user 1. These bulk attenuation corrections were applied to the measured data, to give "synthetic" measurements for the seven pose angles for user 2 that were not measured directly. The result of these measurements is a data set that can be used to characterize the attenuation between iOS devices as a function of many different environmental parameters. The data set is available on GitHub [18].

**Detector Methodology**

The objective of a TCFTL detector is to develop a binary function mapping RSSI estimates into a "yes" or "no" decision regarding the "too close" portion of the detector. Upon analyzing the recorded data, it is obvious that conditioning this decision on the state of the propagation channel (other than the range, which we assume cannot be known a-priori) is advantageous. We believe that some of the channel state information, such as the phone carriage status, may be able to be estimated by the phones themselves using auxiliary sensors such as the light sensor, proximity sensor, and accelerometers, all of which would have little implications on privacy. We believe that it will be more difficult to estimate things like the relative pose angles of the persons carrying the phones, the relative angles of the antennas to each other, and the local multipath environment. As such, we will treat the realizations of these states as random processes, which cause the RSSI measurements to appear stochastic. Thus, we characterize the channel attenuation in terms of a probability density function (PDF), indexed by range of the link and phone carriage status, defined as $p(X/s,c)$, where $X$ is the observed (potentially multi-sampled) data, $s$ is the separation between phones and $c$ is the carriage state of the phones.

For a binary test, such as that in the TCFTL detector, optimality is achieved using a *likelihood ratio test* [19]. Here, optimality can be defined using either the Neyman-Pearson or Bayes criteria, where Bayes criteria requires assigning costs for misclassification of TCFTL. Let us define the state of being "too close for too long" as hypothesis $H_1$, and the state of *not* being too close for too long as $H_0$. Then the likelihood ratio is defined as:

$$L(X,c) = \frac{p(X/H_1,c)}{p(X/H_0,c)},$$

or equivalently, a log-likelihood ratio:

$$l(X,c) = \ln\big(p(X/H_1,c)\big) - \ln\big(p(X/H_0,c)\big).$$

In the case of multiple independent observations $X = x_1 \ldots x_N$, then the ensemble log-likelihoods $\ln(p(X/H_1,c))$ and $\ln(p(X/H_0,c))$ become products of single-sample likelihood terms, and $l(X,c)$ can be expressed as:

$$l(X,c) = \sum_i [\ln\big(p(x_i/H_1,c)\big) - \ln\big(p(x_i/H_0,c)\big)]$$

Or equivalently,

$$l(X,c) = \sum_i z_c(x_i)$$

where $z_c(*)$ is a potentially non-linear transformation of the input data (conditioned on the carriage state). In the Neyman-Pearson optimal test, this likelihood ratio is compared to a threshold; if it exceeds the threshold, event $H_1$ is declared; if it does not exceed the threshold, then $H_0$ is declared. A block diagram can be seen in Figure 2.

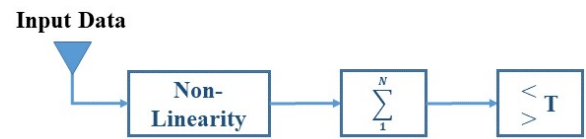

**Figure 2: Block diagram for optimal likelihood ratio binary hypothesis test**

When the PDF of both $p(X/H_1,c)$ and $p(X/H_0,c)$ are Gaussian, the transformation is linear, and the log-likelihood ratio is simply the sum of the data points. If the distribution of the data is log-Normal, then the optimal non-linearity is a logarithmic function; i.e., summing the power of the samples in a dB scale. For the TCFTL detector, we are presented with data that has already been transformed by a logarithmic non-linearity, and is in a dB scale.

Our method for defining the optimization is to consider the true metric for which we would like to optimize for the contact tracing problem. We would like an automatic contact tracing system for which the likelihood of detecting ALL the contacts that are "too close" is possible. This puts a bound on the lowest power we can set the BLE advertisement to ensure it is receivable between phones. At this power level, there will be a maximum range for which our receiver has sufficient sensitivity to correctly decode the BLE message. Ideally, we would like a detector that correctly identifies cases where people are "too close" at a high probability, while minimizing the number of "too close" declarations of phone pairs that exceed 6 feet (phones that are not "too close"). We form a weighted average of the separation-conditioned PDF functions, where the weighting function of separation distance is defined by the physical density of potential contacts as a function of separation distance. Define this physical density as a function of separation from the index case $s$ as $D(s)$. As an example, if we assume that potential contacts are uniformly distributed in 2-D space around the index case, then $D(s)$ will be linear in $s$. Then, the $H_1$ distribution will be:

$$p(X/H_1,c) = \int_0^{\text{Max TCFTL true}} D(s) * p(X/s,c)ds.$$

Equivalently,

$$p(X/H_0,c) = \int_{\text{Max TCFTL true}}^{\text{Max BLE range}} D(s) * p(X/s,c)ds.$$

Using these distributions, we can then generate an optimal likelihood ratio test. Note that for $p(X/H_1,c)$ and $p(X/H_0,c)$ to be valid probability density functions, appropriate scaling of $D(s)$ must be used, and will in general be different for the $H_0$ and $H_1$ cases.

Example PDF estimates from measured data are shown in Figure 1, indexed by inter-phone range ("too close" or "too far") at a fixed phone carriage state, but varying multipath and pose angles. The measured data clearly indicates a trend towards higher attenuation values at longer ranges, but it is also clear that there are large overlaps of PDFs for different ranges. There is in excess of 10 dB path loss difference between the two phone carriage states for a given distance state. Note that for the remainder of this paper, we are using data from iPhones and adjusting the RSSI values to be consistent with a +12dBm input power to the transmitting antenna.

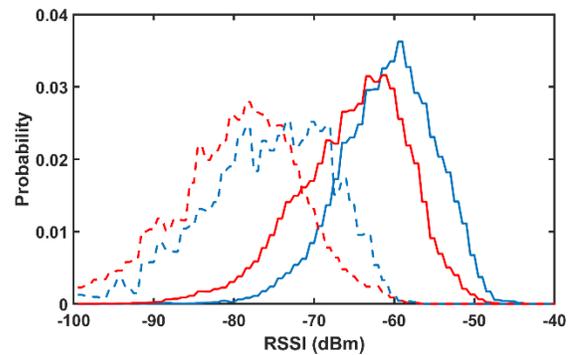

**Figure 1: Estimated RSSI PDFs: Dashed plots correspond to both phones in hand, solid plots correspond to both phones in front pants pockets. Red plots are "too far" cases, blue plots are "too close" cases.**

Given the measured conditional distribution functions of the RSSI data (measured in dB) we have evaluated optimal nonlinearities for different phone carriage conditions as well as different optimization criteria. Figure 3 shows a calculated optimal non-linearity for the case of both phones being in front pants pockets, with $H_0$ corresponding to the weighted PDF for longer separation between phones and $H_1$ corresponding to 6 foot separation:

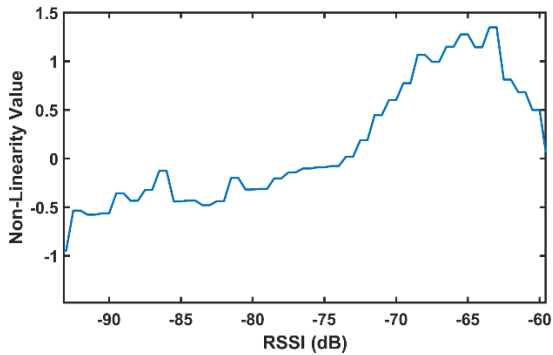

**Figure 3: Estimated optimal nonlinearity for TCFTL detector in case of both phones in pants pockets**

In general, the calculation of this non-linearity is difficult to do accurately because towards the high RSSI levels, the estimated likelihood value for the $H_0$ distribution goes to 0, while the value of $H_1$ may still be relatively large. This generates an ill-conditioned ratio. We compensate for this by applying an arbitrary "floor" for the value of $p(X/H_0,c)$, which sets the maximum value of the calculated non-linearity. Two things become immediately evident: the function is highly non-linear, and data weighting (the magnitude of the weighting coefficients) is very low for low RSSI values, indicating that the optimal detector statistic will be heavily dominated by high-RSSI values. Intuitively, this makes sense, as at these RSSI levels, it is highly unlikely that any $H_0$ event would have occurred. This motivates considering an alternate detector, referred to as the M-of-N detector. In this detector, operating on N samples, a step-function non-linearity is defined (0 below a given threshold, 1 above the threshold), and the likelihood ratio statistic is now simply the count of detections that exceed the threshold. If this count exceeds M, then a detection is declared. These M-of-N detectors are often used in radar signal processing because of their simplicity of operation, and because in many cases, the loss in performance versus an optimal detector, even in Gaussian noise, is quite small. In addition, the detection error tradeoff (DET) curve, a measure of performance of these detectors, can be easily calculated given a measured (non-analytic) PDF. These detectors can be optimized over two coefficients for a given number of samples N; the threshold point τ for the initial nonlinearity, and the value of M chosen. A block diagram for the M of N detector is shown in Figure 4. By the block diagram, it can be seen that the M of N detector can be an optimal likelihood ratio detector for some specific distributions for $H_1$ and $H_0$. Given the measured optimal non-linearity for the TCFTL detector, it is believed that this detector should be close to optimal (in a Neyman-Pearson sense) in performance, while being more simple to optimize for a given detection performance point.

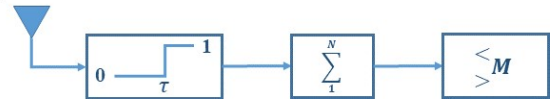

**Figure 4: Block diagram for M of N detector**

For the remainder of this paper, we will consider the performance of these M-of-N detectors, including a further sub-optimal subset of these detectors where M is forced to be 1 (a maximum value detector).

**Data Analysis**

Data was collected and PDFs were generated over a wide range of phone separations and carriage states. From this data, we were able to optimize M-of-N detectors for different optimization criteria and phone carriage states. Note again that for any number of samples N the two detector parameters which require optimization are only the initial RSSI threshold and the count M. For each carriage state, we evaluated detector performance as a function of the number of looks N available to the detector. DET (detection error tradeoff) curves (a way of visualizing the probability of detection versus the probability of false alarm) were calculated by varying the nonlinearity threshold and then minimizing false alarms by optimizing the M value. In the case of the maximum value detector, only the nonlinearity threshold was varied. As an example, the DET curves for choosing N equal to 6, 12, 24, and 48 independent samples when both phones are in pockets are plotted in Figure 5. For reference, we have included the "coin flip" decision curve, which would be achieved if there was no information in the RSSI measurements.

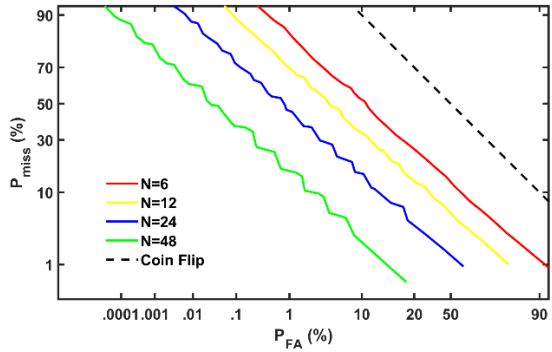

**Figure 5: DET curves for M-of-N TCFTL detector for both phones in pants pockets for various values of N**

We also calculated the DET curves for the same scenario, but this time using only a maximum value detector. This is shown in figure 6.

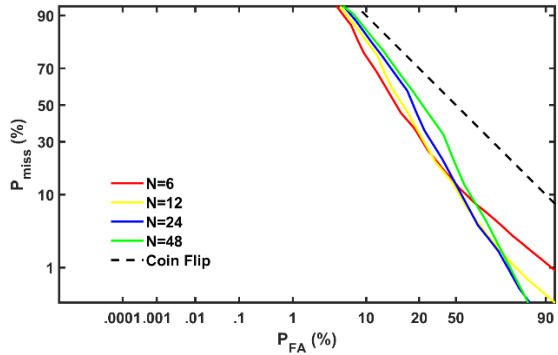

**Figure 6: DET curves for 1-of-N TCFTL detector for both phones in pants pockets for various values of N**

Some key items can be seen in these two plots. As the value of independent samples N increases, the performance of the M-of-N detectors increase at a much higher rate than the performance of the maximum value detector. A second point to note is that for low values of N, the maximum value detector and the M-of-N detector perform almost identically at high values of $P_D$. This can be explained by noting that for small N, it is likely that the optimal value of M is 1.

To illustrate how performance varies with phone carriage, we also show in Figures 7 and 8 the DET curves for the case where body shadowing is lessended, when phones are carried in shirt pockets.

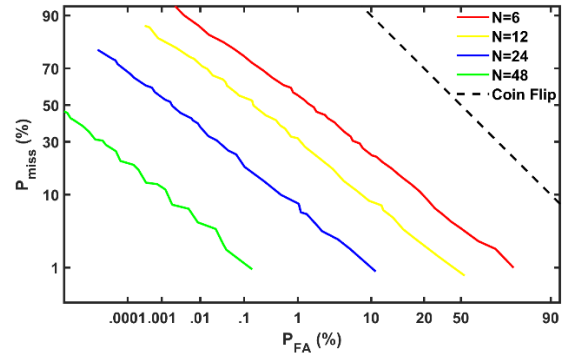

**Figure 7: DET curves for M-of-N TCFTL detector for both phones in shirt pockets for various values of N**

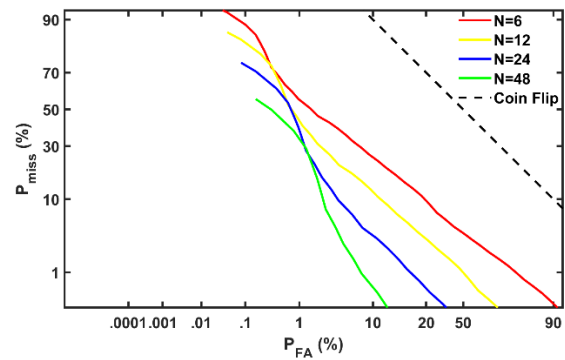

**Figure 8: DET curves for 1-of-N TCFTL detector for both phones in shirt pockets for various values of N**

Note that the overall detector performance in this case is much better, because the spread of shadowing loss tends to decrease as the phone is extended from the body. In the case of shirt pockets, the phones tend to be further from the body than when they are in pants pockets. The trend where M-of-N detectors outperform maximum value detectors for higher values of N continues.

It is also instructive to look at the optimal nonlinearity for this phone carriage state. Figure 9 shows that while the structure of the nonlinearity remains similar to that for pants-pocket carriage, the point at which the weight dramatically increases shifts to higher RSSI values. This implies that the threshold which provides a given $P_D$ for M-of-N detectors or maximum value detectors will change dramatically based on phone carriage state.

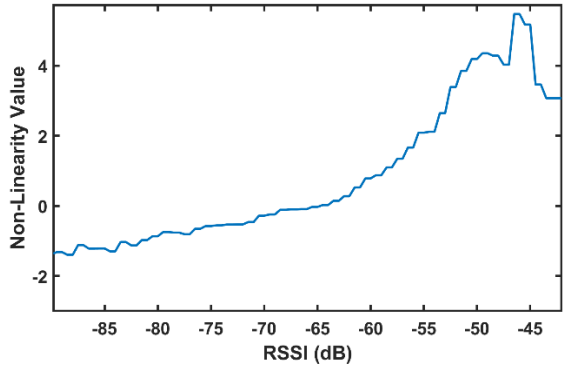

**Figure 9: Estimated optimal nonlinearity for TCFTL detector in case of both phones in shirt pockets**

It is quite possible that the phone carriage state will change during the "too long" period of observation used for the TCFTL detector. Up to this point, we have assumed that we can use an optimal M-of-N detector for each specific phone carriage state – but the true detector must operate when some samples are drawn from one phone carriage state and other samples are drawn from different phone carriage states. Our proposed method of implementation for this detector is as follows:

- On a sample-by-sample basis, correct the measured RSSI value based on the known phone carriage state of that measurement. This corresponds to a fixed addition/subtraction to the RSSI value unique to the particular phone carriage state, such that each state would now have aligned nonlinearity thresholds for their optimal detectors
- The M-value and thresholds will be calculated using a minimax strategy. Given that the realization of phone carriage cannot be controlled, we pick the nonlinearity threshold and M value to give the best performance possible under the worst-case phone carriage state. This worst-case state can be calculated by our previous single-phone-carriage state detector optimizations. The resulting detector will therefore have the best worst-case performance over any realized set of phone carriage conditions.

If phone carriage information is not available to the detector, it is clear from these plots that using a common M-of-N or maximum value detector across all phone carriage states will provide sub-optimal performance, as in some carriage situations the thresholds will be too low, while in other carriage situations, the thresholds will be too high. To understand the implications of different detector performance points, we introduce the concept of False Discovery Rate (FDR). FDR attempts to capture the ratio of false contact declarations (declaring a contact as being too close for too long, when in fact it was not too close) to the total number of contact declarations. If this ratio equals 1, then all declared contacts would be false. If the ratio is 0, then all of the declarations correspond to true detections. This metric is important in automated contact tracing, because for the contact tracing system to be effective, a declaration of TCFTL should initiate a series of events, potentially including self-quarantine and the contact of local public health authorities, who may schedule testing. A high FDR will unnecessarily initiate this sequence of events in a potentially large group of people. In addition, if the FDR of a system is known to be high, public confidence in the efficacy of the app may be eroded, negating any benefit to overall public health due to people ignoring TCFTL declarations made by the app.

In order to evaluate FDR, we first have to hypothesize prior distributions on both the physical distribution of potential contacts around an index case (similar to the way we did when generating $H_0$ and $H_1$) and the phone carriage states of the contacts and index case. For a given phone carriage state, we can calculate the expected number of true contacts ($TC$) in the following manner:

$$TC = \int_0^{\text{Max TCFTL true}} D(s) * P_D(s) ds,$$

where $D(s)$ is defined as before (but in this case, remains unscaled), and $P_D(s)$ is the probability of declaring detection of a contact for a given detector at range $s$. Similarly, the expected number of false contacts $FC$ for a given phone carriage will then be given by:

$$FC = \int_{\text{Max TCFTL true}}^{\text{Max BLE range}} D(s) * P_D(s) ds.$$

Note that the integrand is the same for both $FC$ and $TC$. In $FC$, the expression $P_D(s)$ would be referred to as the probability of false alarm (since the argument $s$ is confined to ranges at which there are no true contacts to detect), but it is in fact the same function as that used in calculating $TC$. We can take a weighted sum of these expected values (where the weights correspond to the assumed probabilities of a given phone carriage state) to generate the overall expected number of true contact declarations and false contact declarations for a given assumed density of contacts. For the remainder of the paper, we will assume that the potential contacts are distributed uniformly in 2-D space. We will vary our assumptions on phone carriage state to evaluate how that effects outcomes. Note that for ease of calculation, we will continue to assume that for a given TCFTL test, the phone carriage states will remain constant for the duration of the contact.

One issue faced in our calculation of an overall FDR is that, due to measurement limitations, we only have data sets collected to inter-phone ranges of 15 feet. However, we realize that propagation, even at lower power levels, will allow reception of BLE signals at up to 30 feet. To solve the problem of lack of data, we have made some approximations. Our first approximation is that at ranges of 15 feet or longer, it is likely that the local multipath environments around the two phones are somewhat decoupled – thus the effects of multipath on the distribution of RSSI values will effectively be the same for all phone separations in excess of 15 feet. We also assume that the shadowing effects due to pose angle of the phone carriers will effectively become independent of range. Under these assumptions, we can then take the measured RSSI data at 15 foot phone separation and simply "shift" the values by an amount appropriate for the excess path loss commensurate with a longer inter-phone separation, and generate PDFs based on this synthesized data, which then can be used in assessing detection probabilities at these extended ranges.

Based on this methodology, we look at the performance of the baseline Apple|Google (A|G) API, where we consider two prior distributions for phone carriage:

- Equally likely that any of the phone carriage conditions for which we have current measurements occur
- Phone carriage biased such that there is a 30% chance of carrying in hand, and 40% probability of being in the standing position

If results for these two different priors are similar, we feel that the results will be somewhat independent of actual phone carriage statistics.

The following table illustrates the limited phone carriage cases for which data was available for this analysis:

| User 1 | User 2 |
|---|---|
| Standing, phone in bag | Sitting, phone in hand |
| Standing, phone in front shirt pocket | Standing, phone in hand |
| Standing, phone in front pants pocket | Sitting, phone in front shirt pocket |
| Standing, phone in hand | Standing, phone in front pants pocket |
| Standing, phone in front pants pocket | Standing, phone in front pants pocket |
| Sitting, phone in hand | Sitting, phone in hand |

**Table 1. Phone states of the analyzed data sets.**

Given that there have been multiple iterations of options available for contact tracing scoring in the A|G API, we need to state our understanding of the current state of affairs. Each phone with the A|G API will constantly produce privacy-protecting "chirps" at a roughly 4 Hz rate. Each phone will scan for these chirps no less frequently than once every 5 minutes, with a scan time of 4 seconds. However, the expected number of scans per 15 minute period is closer to 6, given the ability of the phones to opportunistically scan for BLE chirps. For each uniquely identified contact, the receiver will record the time of arrival of the chirp, the RSSI the chirp was received at, and a transmitted power level encoded into the chirp that represents the transmitting phone's estimate of the power put into its antenna. When a generating key corresponding to a known index case is presented to

the phone, the receiving phone will compare the set of pseudo-random bit sequences that the key generates to all of the chirp bit sequences it has recorded for some predetermined amount of time in the past. If the key-generated sequences match any received chirp sequences, then the API will present the public-health authorized automated contact tracing app some subset of the recorded data. These data will at least contain the number of chirps received with attenuations below an app-defined threshold, and the minimum attenuation level estimated from that emitter. Thus, an M of N detector is possible, as well as a maximum value (minimum attenuation) detector.

There is some uncertainty as to what data will be retained in each 4-second receiver scan, during which time it is likely that multiple chirps from any nearby device will be heard. For completeness, we consider three cases: only the first chirp heard from any given device during a scan period is recorded; all chirps from a given device are recorded during the scan period; and finally, only the chirp from a given device with the minimum attenuation during the scan period is recorded. The performance under these different assumptions will differ. The performance of the detector where the minimum attenuation value is provided per scan would result in a 1-of-N detector, which will, by definition, be no better than an M-of-N detector for any value of N. Thus, for larger values of N we will plot only the M-of-N results, which will serve as an upper bound on the performance advantage which can be gained by using multiple samples per scan period.

Under the first assumption, that only the first chirp received from a given device during a scan is recorded, performance for the expected number of chirps that would be seen within a "too long" period is given in Figure 10.

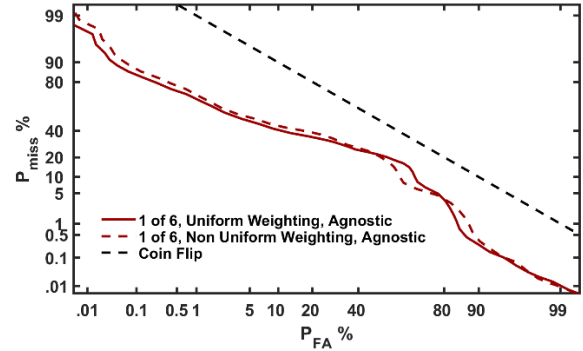

**Figure 10: DET curves for phone-carriage agnostic detector using 6 looks. Weighting refers to fraction of time a phone is assumed to be carried in any one state.**

It is evident that the performance of this detector will not be heavily dependent on our a-priori assumption of phone carriage distribution. It is also obvious from these plots that the performance of the state-agnostic detector will be poor; at a probability of detection of 0.7 ($P_{miss}$ = 30%), the probability of false alarm is about 40%. This is because the thresholds used for the detector must be fixed for all phone carriage situations. However, the distributions for $H_1$ and $H_0$ are highly contingent on phone carriage state, as could be seen in Figure 1. Choosing a single threshold to separate between $H_1$ and $H_0$ with low false alarms and high probability of detection for both of the phone carriage states in Figure 1 would be impossible – the $P_{FA}$ for phone in hand would exceed the $P_D$ for phone in pants pocket.

Performance may improve if we assume that during each 4-second scan period, we can record all of the RSSIs seen for a given index case and use them for the detection statistic. Note that it is assumed that the RSSI values within a four-second window will appear correlated, as pose angle and multipath conditions are unlikely to change dramatically within that time frame. To emulate this situation, we chose 4 samples for a given pose angle at a given range, corresponding to being able to take four looks within one scan period.

Given the expected performance of the baseline detector, we have also evaluated the performance of some modified versions of the basic detector, including the usage of phone carriage states in setting individual thresholds for attenuation estimates prior to M-of-N testing. We have considered these state cognitive detectors using different look values for N. In using this "cognitive" detector (i.e., cognitive of phone carriage state), we have optimized the individual thresholds for the different phone carriage states such that for any given overall average $P_D$, we pick the set of thresholds that minimizes the resulting overall average $P_{FA}$. As an example, for an overall $P_D$ of 0.6, the resulting spread in thresholds spans 20 dB.

We now show in Figure 11 the performance of both phone carriage state cognitive and phone carriage state agnostic detectors for the 6 independent and 24 correlated sample cases:

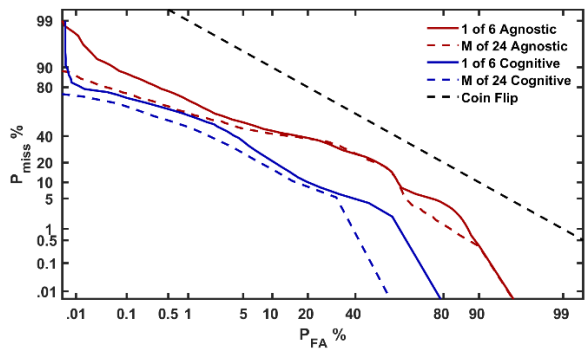

**Figure 11: DET curves for 6 independent and 24 correlated looks, for phone carriage state agnostic and phone carriage state cognitive detectors**

It is clear that both phone carriage state cognizance and additional correlated looks are important factors in the overall performance of these detectors. We can now look at how these detectors will perform in the FDR sense. Since FDR will be a function of the probability of detection at which the detector is chosen to operate, we plot FDR versus $P_D$ in Figure 12.

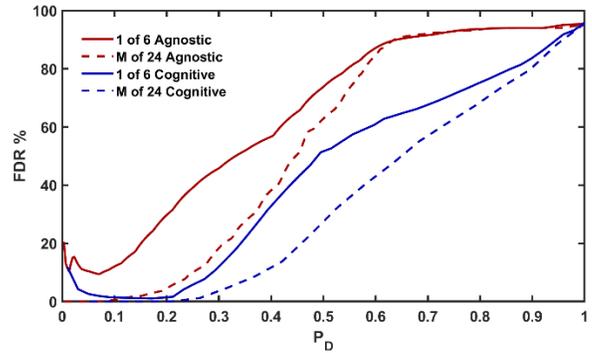

**Figure 12: FDR percentage for detectors as a function of $P_D$ – 6 uncorrelated samples and 24 correlated samples**

Again, the benefit of both state cognition and increased numbers of correlated looks is shown clearly. Consider some potential operating points for a detector. In one case, we assume that we want a very high probability of detection (say, 80%). For the either the baseline 6-sample agnostic detector or the augmented 24 sample correlated agnostic detector, the FDR would be close to 95%. That means that the number of false declarations per "true" TCFTL events would be about 20:1. If, however, we use the cognitive detector with 6 samples, the FDR has decreased to less than 80%. This would produce about four false declarations for each "true" TCFTL event. If we use the phone carriage cognizant detector with 24 correlated looks, then the FDR decreases further to around 66%, which corresponds to roughly 2 false declarations per "true" TCFTL event.

Another operating point would be to pick an acceptable FDR rate, and observe what $P_D$ would be achievable. For the baseline 6-sample agnostic detector, to achieve an FDR of 50% (where there is no more than one false declaration per true declaration), the $P_D$ must be lower than 0.3. Even allowing the use of 24 correlated samples only increases the $P_D$ to about 0.40. However, if we allow the use of a cognitive detector, the $P_D$ ranges from 0.5 with 6 uncorrelated looks to greater than 0.6 for 24 correlated looks. Thus, at an FDR of 50%, the cognitive detector would be 67% more effective with only 6 looks and 50% more effective with 24 correlated looks than the comparable agnostic detectors.

Finally, we look in Figure 13 at what benefit additional looks, both correlated and uncorrelated, would provide the phone state cognitive detectors. Here, we look at the effects in terms of $P_D$ for a fixed FDR of 50%. It is clear that having uncorrelated samples vs. correlated samples is a large benefit; it also becomes clear that at this high $P_D$ level extra correlated samples beyond about 25 are of clearly providing diminishing returns. This means that taking more than 4 individual samples in each 4-second dwell will be unlikely to improve performance. However, taking 10 scans (10 independent looks) with only one look per scan has superior performance than taking an arbitrarily large number of looks within each of 6 scans.

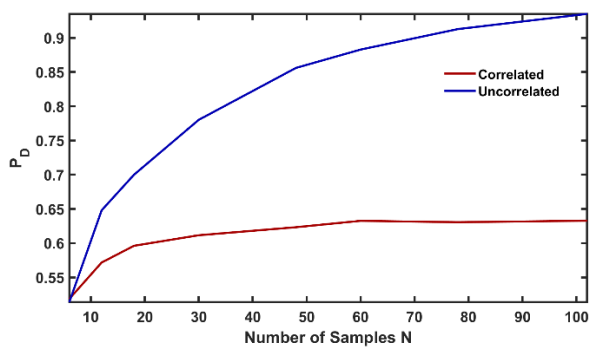

**Figure 13:** $P_D$ percentage for state cognitive detectors as a function of number of samples – correlated samples taken within 6 BLE scans over 15 minutes and uncorrelated samples (increased BLE scans per 15 minutes), both with FDR = 50%.

**Summary**

Contact tracing is a proven method of mitigating the spread of infectious diseases. Automated contact tracing could be a strong augmentation to manual contact tracing, but in order to achieve this potential, a method for accurately identifying higher risk contacts of an infectious "index case" automatically must be developed. Bluetooth Low Energy (BLE) beacons have been proposed as a method for accomplishing this detector, but it has many impediments to accurately distinguishing between contacts that are truly "too close" (as defined by the CDC) from those that are actually "too far".

Understanding the performance of these detectors is important for public health authorities in their determining the efficacy and effectiveness of automated exposure notification as part of their response to the covid-19 pandemic. This paper examines methods of building detectors from BLE Received Signal Strength Indication (RSSI) measurements. Our results indicate that the baseline (as currently understood) performance of the Apple/Google-API-enabled detectors will have performance that allows detection of roughly 30% of true "too close" events if we limit the number of misclassified "too far" events to be equal to the number of truly "too close" events detected. This is indicated in the false discovery rate (FDR) vs. detection probability curves for the agnostic detectors in Figure 12. However, with some feasible augmentations, we believe a detector with superior performance (greater than 60% correct identification of true TCFTL events, with an FDR of 50%) can be constructed. Specifically, an application of M of N detectors has been proposed which can provide robust performance when coupled with potentially available auxiliary data. We have collected data sets to allow us to evaluate the performance of these detectors, and we have shown the importance of using detectors which utilize both potentially available auxiliary data and signal strength sampling rates that provide between 20 and 30 (potentially correlated) samples per "too long" period. We also show that increasing the number of scan periods within the "too long" time period can have further positive effects on detector performance, as shown in Figure 13.